# Highlights

## Tidal dissipation of binaries in asteroid pairs

Laurent Pou,Francis Nimmo

- We study 13 binary asteroid systems inside asteroids pairs with tides and thermal torques.
- All secondaries are found to be consistent with rubble piles.
- The $Q/k_2$ ratio of (3749) Balam 1 is constrained to be between 2.7 $\times 10^4$ and 1.4 $\times 10^6$

# Tidal dissipation of binaries in asteroid pairs

Laurent Pou[a,b,*], Francis Nimmo[a]

[a]*University of California Santa Cruz, Dept. Earth and Planetary Sciences, 1156 High Street, Santa Cruz, 95064, CA, United States*
[b]*Jet Propulsion Laboratory, California Institute of Technology, 4800 Oak Grode Drive, Pasadena, 91109, CA, United States*

ABSTRACT

Tidal dissipation in a celestial body can be used to probe its internal structure. Tides govern the orbital evolution of binary systems and therefore constraints on the interior of binary system members can be derived by knowing the age and tidal state of the binary system. For asteroids, age estimates are challenging due to a lack of direct observation of their surface. However, the age of asteroid pairs formed by rotational fission of a parent body can be derived from dynamical modeling, and as such can be used to constrain the age of binary systems existing within asteroid pairs. We study 13 binary asteroid systems existing in asteroid pairs by modeling their tidal locking and eccentricity damping timescales from tidal dissipation in the primaries and secondaries. We consider the impact of thermal torques on these timescales from the YORP and BYORP effects. The resulting constraints on the tidal dissipation ratio $Q/k_2$ are compared to monolithic and rubble pile asteroid theories, showing that all secondaries are consistent with rubble piles with regolith layers greater than 3m and suggest that $Q/k_2$ for rubble piles increases with radius. A particular case is the first bound secondary of asteroid (3749) Balam, whose $Q/k_2$ is constrained to be between $2.7 \times 10^4$ and $1.4 \times 10^6$, consistent with a rubble-pile with a regolith thickness between 30m and 100m.

## 1. Introduction

Many inner belt asteroids are part of binary or multiple systems, with the binary fraction among asteroids smaller than 15 km in the inner belt estimated to be at least 11% (Margot et al., 2002; Pravec et al., 2002). For such bodies, tidal evolution theory can be used to estimate their internal properties (e.g., Taylor and Margot, 2011) by identifying the tidal end states of binary systems, and comparing them to the dynamical properties of the currently observed systems. Of particular interest is the derivation of the tidal dissipation quantity $Q/k_2$ where $Q$ is the quality factor related to the phase lag of the tidal response of the body (e.g., Efroimsky and Makarov (2013)) and $k_2$ is the Love number quantifying the gravity response of the body due to tides (e.g., Love, 1911; Munk and MacDonald, 1960). The spin, semi-major axis, and eccentricity evolution of system are governed by this ratio which depends of the amplitude and phase of an object's response to tides (e.g., Goldreich and Gold, 1963; Meyer et al., 2023), and can be used to infer their internal structure (e.g., Moore and Schubert, 2000).

When trying to constrain asteroid tidal dissipation, a major challenge is to determine the time over which tidal evolution has taken place. When direct observation of the surface to estimate the age of an object (e.g., Yoder, 1982, for Phobos) is not possible, typical assumptions for this timescale are the age of the Solar System (Margot and Brown, 2003; Taylor and Margot, 2011), or the dynamical lifetime expectancy of asteroids before they get flung into Jupiter-crossing orbits or towards the Sun (Gladman et al., 1997; Margot et al., 2002).

Asteroid pairs are pairs of genetically related asteroids on highly similar heliocentric orbits (Pravec et al., 2019). Sometimes, one member of the pair is actually a binary system, a situation we will refer to as "binary systems in asteroid pairs". The advantage of pairs is that, with rotational fission being the most likely cause for the formation of these pairs (Scheeres, 2007; Pravec et al., 2010, 2019), the time since separation between the two members of each asteroid pair can be estimated and thereafter used to constrain the actual age of the binary system inside the asteroid pair as an upper bound.

We study here the tidal evolution of binary asteroid systems with a focus on the spin rate evolution of both bodies and the evolution of the eccentricity of the secondary of the binary system. In Section 2, we derive the tidal end states of a binary asteroid systems and their link with the tidal dissipation ratio $Q/k_2$ of asteroids. Thermal torques arising from the YORP effect (Rubincam, 2000) and the BYORP effect (Ćuk and Burns, 2005) are known to impact the orbital evolution of asteroids and are estimated in Section 3 following the approaches of Goldreich and Sari (2009) and Jacobson and Scheeres (2011b). We apply these results on binary systems in asteroid pairs detected in Pravec et al. (2019) in Section 4 and 5 to get upper and lower bounds on their tidal dissipation, and compare the resulting values to the previous works of Goldreich and Sari (2009), Jacobson and Scheeres (2011b), and Nimmo and Matsuyama (2019). Finally, in Section 6 we discuss the conclusions and expand on neglected effects in our study such as the impact of primary tides on eccentricity (Murray and Dermott, 2000; Jacobson and Scheeres, 2011b), higher-order tides (Taylor and Margot, 2010), and high initial eccentricity Boué and Efroimsky (2019).

*Corresponding author
✉ laurent.pou@jpl.nasa.gov (L. Pou)
ORCID(s):

## 2. Tidal evolution of binary asteroids
### 2.1. Tidal locking

We consider a binary asteroid system made of a secondary asteroid orbiting a larger, primary asteroid. We first focus on the primary and consider the tide raised on the primary by the secondary moving on a circular, equatorial orbit of mean motion $n$ and semi-major axis $a$, while the spin of the primary is denoted $\Omega_p$. In all this paper, all terms in all equations are given in SI units. Physical parameters with a $p$ subscript (such as $\Omega_p$) refer to the parameter for the primary, while $s$ subscripts refer to the secondary body.

Following Murray and Dermott (2000), when considering only degree 2 tides from the secondary deforming the primary from a sphere into a near-spherical body, the semi-major axis $a$ evolution can be written as:

$$\dot{a} = sign(\Omega_p - n) \frac{3k_{2p}}{Q_p} \frac{m_s}{m_p} \left(\frac{R_p}{a}\right)^5 na \qquad (1)$$

where $k_{2p}$ is the Love number $k$ for degree-2 tides linked to the primary, $Q_p$ is the tidal dissipation factor of the primary, $m_p$ and $m_s$ are the mass of the primary and the secondary, respectively, and $R_p$ is the mean radius of the primary. Here we are assuming that there is no dissipation in the secondary.

Assuming conservation of angular momentum in the system, the change of the primary spin is given by:

$$\dot{\Omega}_p = -sign(\Omega_p - n) \frac{3k_{2p}}{2\alpha_p Q_p} \frac{m_s^2}{m_p(m_p + m_s)} \left(\frac{R_p}{a}\right)^3 n^2 \qquad (2)$$

where $\alpha_p$ is the coefficient in the moment of inertia of the primary along its main axis $I_p$ such as $I_p = \alpha_p m_p R_p^2$, with $\alpha_p = \frac{2}{5}$ for a uniform sphere, and less otherwise.

Using the same arguments for the tide raised on the secondary by the primary, the spin of the secondary $\Omega_s$ evolves according to the following equation:

$$\dot{\Omega}_s = -sign(\Omega_s - n) \frac{3k_{2s}}{2\alpha_s Q_s} \frac{m_p^2}{m_s(m_s + m_p)} \left(\frac{R_s}{a}\right)^3 n^2 \qquad (3)$$

Because of the term $-sign(\Omega - n)$, we can see that tidal evolution will drive the spin $\Omega$ of both bodies towards the mean motion $n$ in a state called tidal locking. This state will be reached earlier for dissipative bodies (meaning a high $k_2/Q$ or a low $Q$) and for the secondary as its mass is lower than the mass of the primary.

### 2.2. Eccentricity damping

In the case of an eccentric orbit, tidal dissipation can also cause the eccentricity $e$ to change. For now, we will assume eccentricity to be small and focus on tidal dissipation in the secondary; tidal dissipation in the primary and its impact on eccentricity is neglected here and will be discussed in section 6.1.

As given in Murray and Dermott (2000), assuming once again conservation of angular momentum, the change of eccentricity $\dot{e}$ due to tides in the secondary is:

$$\dot{e} = -\frac{21}{2} \frac{k_{2s}}{Q_s} \left(\frac{R_s}{a}\right)^5 \frac{m_p}{m_s} ne \qquad (4)$$

Thus, tides in the secondary always act to decrease the orbital eccentricity until the secondary's orbit is circular, at which point the eccentricity $e = 0$.

### 2.3. Tidal timescales

By using Eq. 2, Eq. 3, and Eq. 4, we calculate the tidal locking timescales $\tau_\Omega$ and eccentricity damping timescale $\tau_e$:

$$\tau_{\Omega_p} = \left|\frac{\Omega_p}{\dot{\Omega}_p}\right| = \frac{2\alpha_p}{3} \frac{Q_p}{k_{2p}} \frac{m_p(m_p + m_s)}{m_s^2} \left(\frac{a}{R_p}\right)^3 \frac{\Omega_{init}}{n^2} \qquad (5)$$

$$\tau_{\Omega_s} = \left|\frac{\Omega_s}{\dot{\Omega}_s}\right| = \frac{2\alpha_s}{3} \frac{Q_s}{k_{2s}} \frac{m_s(m_s + m_p)}{m_p^2} \left(\frac{a}{R_s}\right)^3 \frac{\Omega_{init}}{n^2} \qquad (6)$$

$$\tau_e = -\frac{e}{\dot{e}} = \frac{2}{21} \frac{Q_s}{k_{2s}} \frac{m_s}{m_p} \left(\frac{a}{R_s}\right)^5 \frac{1}{n} \qquad (7)$$

where $\Omega_{init}$ is the initial spin of each body, arbitrarily chosen to be the primary's current spin for the secondary timescale and twice as fast for the primary timescale to provide a robust upper bound on $Q/k_2$ for tidally locked bodies.

Due to the mass and size difference between the secondary and the primary, tidal locking of the secondary typically occurs before its orbital eccentricity is fully damped. Knowledge of the current state of a binary system can be converted into constraints on the interior of the asteroids using these timescales. For instance, if the secondary is tidally locked, its age must be greater than the tidal locking timescale for the secondary $\tau_{\Omega_s}$. Hence, if the age of the binary system and its state are known, it is possible to constrain the tidal dissipation rate $Q/k_2$ of the system. However, this is only true if only tides affect the evolution of the spin and eccentricity of the system. When studying small bodies such as asteroids, this is not true as thermal torques from radiative forces must also be considered.

## 3. Radiative forces and orbital evolution

Under radiation from the Sun, asteroids experience a thermal drag due to the Yarkovsky effect (Rubincam, 1995). Asteroids first absorb the sunlight before reradiating it in the infrared; due to the momentum carried by the reradiated photons, this creates a net force on the asteroid capable of altering its semi-major axis over large timescales (e.g., Vokrouhlický et al., 2000; Žižka and Vokrouhlický, 2011). This force impacts mostly small objects of less than 40 km in diameter (Vokrouhlicky et al., 2015) and thus is usually neglected for planetary evolution, but must be considered for asteroid studies.

More relevant to our work is the rotational counterpart of the Yarkovsky effect called the Yarkovsky–O'Keefe–Radzievskii–Paddack effect, or YORP effect (Rubincam, 2000). The YORP effect stems from the irregular shape of the asteroid, where the reradiated photons are reemitted in different directions causing a net torque on the asteroid, altering its spin. This effect was first measured on asteroid (54509) 2000 PH5, now renamed (54509) YORP (Taylor et al., 2007; Lowry et al., 2007), and is likely to be the origin of the formation of most binary asteroid systems (Scheeres, 2007; Pravec et al., 2010).

For a binary asteroid system, another effect occurs when the secondary reaches tidal locking. As its spin matches its mean motion, its orientation along its orbit is no longer random and it develops permanent leading and trailing hemispheres. This means that over its mean motion, any shape irregularity of the secondary will create a net torque on the orbit of the secondary in a similar way that the the YORP effect does over a full spin rotation. This effect has been named Binary YORP effect, or BYORP effect (Ćuk and Burns, 2005). Also due to the shape of the secondary, the BYORP effect can alter the semi-major axis and the eccentricity of the secondary's orbit: as such, both YORP and BYORP effects will impact the tidal locking and eccentricity damping timescales calculated before and must be taken into account to fully model the orbital evolution of a binary asteroid systems.

### 3.1. YORP effect

To model the YORP effect, we follow the formalism of Goldreich and Sari (2009): here, the rate of change in spin $\Omega$ for an asteroid is given by:

$$(\dot{\Omega})_{YORP} = \frac{15\sqrt{3}}{64\pi^{3/2}} f_y \frac{L_S m}{c\rho^{3/2} R^{3/2} a_S^2} \qquad (8)$$

where $m$, $\rho$, and $R$ are respectively the mass, density, and radius of the considered body (either primary or secondary in our case), $c$ is the speed of light, $a_S$ is the distance to the Sun, and $L_S$ is the solar luminosity. $f_y$ is a factor quantifying the strength of the YORP effect on the asteroid, mainly impacted by its shape. Based on the measurements of the spin-up rate of asteroid (54509) YORP in Taylor et al. (2007) and Lowry et al. (2007), they found a YORP factor $f_y \approx 4 \times 10^{-4}$.

In our models, we keep a similar value of the YORP factor $f_y$ but we allow it to be either positive or negative. This rate of change in spin is then added to the change of primary spin and secondary spin from Eq. 2 and Eq. 3, respectively, to estimate a range of possible tidal locking timescales depending on the tidal dissipation ratio $Q/k_2$ of the studied body and on the strength of the YORP effect. While the tangential YORP effect (Golubov and Krugly, 2012; Golubov et al., 2014) may also contribute to the spin evolution of asteroids, due to the uncertainties in its strength and implementation, we did not implement it in this study.

### 3.2. BYORP effect

McMahon and Scheeres (2010a,b) showed that the BYORP effect impacts the evolution of the semi-major axis and the eccentricity of the secondary, and that this evolution can be quantified by a single $B$ coefficient called BYORP coefficient representing the averaged acceleration in the direction parallel to the motion of the secondary.

Following Jacobson and Scheeres (2011b), the evolution of eccentricity due to the BYORP effect is:

$$(\dot{e})_{BYORP} = \frac{3}{8\pi} B \frac{F_S}{a_S^2 \sqrt{1-e_S^2}} \frac{e}{\omega_d \rho} \left(\frac{a}{R_p}\right)^{1/2} \frac{1}{R_p^2} \frac{\sqrt{1+q}}{q^{1/3}} \qquad (9)$$

where $F_S$ is the solar radiation constant, $e_S$ is the eccentricity of the heliocentric orbit, $\omega_d = \sqrt{4\pi G \rho/3}$ is the spin disruption limit of a sphere, and $q = m_s/m_p$ is the mass ratio between the secondary and the primary. Based on the maximum value used in Jacobson and Scheeres (2011b), for our modeling we use $B = 10^{-2}$ and allow $B$ to be either negative or positive, meaning the BYORP effect can act either to increase or reduce the eccentricity of the secondary's orbit. In the similar way as the YORP effect, we add the contribution of the BYORP effect to the evolution of the eccentricity due to the secondary's tides from Eq. 4 to get a range of possible eccentricity damping timescales depending on the tidal dissipation ratio $Q_s/k_{2s}$ of the secondary and on the strength of the BYORP effect.

## 4. Asteroid pairs

From equations (5), (6) and (7) it is possible to estimate the time needed for the components of a binary system to reach synchronous rotation and/or a circular orbit depending on their tidal dissipation ratio $Q/k_2$. However, the current orbital elements of the binary system and its age must be known to infer the actual value of the $Q/k_2$ ratio and therefore constrain its interior properties (e.g., Taylor and Margot, 2011). Typical constraints include the age of the Solar System (Margot and Brown, 2003; Taylor and Margot, 2011) or the dynamical lifetime expectancy of asteroids (Gladman et al., 1997; Margot et al., 2002), but are only rough upper bounds on the true age of the system.

When trying to determine the age of asteroids, asteroids pairs are of particular interest. Asteroids pairs are asteroids on highly similar heliocentric orbits (Vokrouhlický and Nesvorný, 2008) which cannot have formed from random coincidences and thus must be genetically related. Studies of their spin and orbital evolution (Pravec et al., 2010) showed that the correlation between the rotation frequencies of asteroid pair primaries and the asteroid pair mass ratios is evidence that they were created through rotational fission (Scheeres, 2007). Backwards orbit integration including modeling the Yarkovsky effect (e.g., Vokrouhlický and Farinella, 1999) acting on each member of asteroid pairs can be done to determine close encounters and therefore estimate the time between separation of the two members of the asteroid pairs as a way to evaluate their age.

Pravec et al. (2019) identified 13 asteroid pairs that were also harboring multiple systems, mostly binaries. As the preferred means of explaining the existence of both the multiple asteroid systems linked to a member of the asteroid pair and the asteroid pairs themselves is rotational fission (Scheeres, 2007), either through secondary fission (Jacobson and Scheeres, 2011b) or cascade fission (Pravec et al., 2018, 2019), this would mean the age of the asteroid pair can be used as an upper bound for either the age of the multiple system, or the last time (neglecting possible impacts) since its spin rate and orbit were chaotically excited before being able to go through tidal locking and eccentricity damping. Coupled with the previously derived timescales, this age constraint provides a way to estimate the $Q/k_2$ of the members of the multiple system existing inside the asteroid pair.

## 5. Application to binaries in asteroid pairs

The list of multiple systems which are themselves part of an asteroid pair and observed properties from Pravec et al. (2019) is given in Table 1. Most binary systems observed have their secondaries tidally locked ($P_s = P_{orb}$) and their eccentricity damped ($e = 0$), while the primaries are not in synchronous rotation. Exceptions are both secondaries of asteroid (3749) Balam, where the first is tidally locked but on an eccentric orbit, and the second is neither tidally locked nor on a circular orbit, as is also the case for the secondary of asteroid (21436) Chaoyichi. The age of asteroid (10123) Fideöja is ill-constrained as no upper bound was obtained in Pravec et al. (2019), making it ill-suited to our tidal and radiative constraints and will thus be omitted in the rest of this paper. We also assume that the primary and secondary have the same bulk density, so that mass ratios in our equations can be directly derived from the observed size ratio.

For each multiple system inside an asteroid pair, we derive the tidal locking and eccentricity damping timescales due to both tides and radiative torque effects for a large range of possible $Q/k_2$ for both the primary and each secondary of the multiple system. This is illustrated for the secondary of (26420) 1999 XL103 in Fig. 1, where time is plotted against the $Q/k_2$ of the secondary. Because of the opposite signs between the radiative effects and tidal effects on the orbital elements, + YORP and + BYORP means that the radiative torques are working against tides, while - YORP and - BYORP means that they are working in the same direction as tides. When tidal dissipation is strong (i.e., low $Q_s/k_{2,s}$), the radiative torques have little effect on either timescale. However, depending on the signs of the YORP and BYORP effects, two possibilities can occur for high $Q_s/k_{2,s}$ values where tidal dissipation is negligible. If the radiative torques works in the same direction as tides, the timescales become horizontal lines, as the radiative effects are dominant and move the system towards the tidal end states at a speed independent of their interior properties. On the other hand, when the radiative effects are opposite to tidal effects and tidal dissipation is low, it is impossible to reach the tidal end states, which is visible in the figure as timescales with + YORP and + BYORP become vertical as the $Q_s/k_{2,s}$ increases i.e., tidal effects get weaker. The derived timescales are then compared to the age of the asteroid pair to which the primary asteroid belongs, and depending on the current state of the binary system, bounds are obtained on $Q/k_2$.

In this particular example, the secondary of (26420) 1999 XL103 is both synchronous and circularized, which must have happened within the upper bound on the pair age, 838 kyr. Because despinning happens more rapidly than circularization, the circularization time provides the upper bound on the secondary $Q/k_2$, which based on Figure 1 is $7.9 \times 10^4$ once BYORP is taken into account.

Table 1: Binary (or more) systems inside asteroid pairs, based on Table 1 and 4 from Pravec et al. (2019). The age given is the age of the asteroid pair the bound primary of the binary system is part of. Error bars on orbital parameters are based on the Binary Asteroid Database from the Ondrejov Asteroid Photometry Project available online at https://www.asu.cas.cz/ asteroid/binastdata.htm (Pravec and Harris, 2007; Pravec et al., 2012, 2016, 2019). Asteroid (3749) Balam is mentioned twice, as it is part of the multiple system with two separate bound secondaries. Bulk densities are assumed to be the same between primaries and secondaries. When the ratio $D_s/D_p$ is ill-defined (marked with *), the minimum value is used in the rest of the paper. Values marked with † are not directly measured but assumed based on comments made in section 5.1 of Pravec et al. (2019) indicating it is likely that most bound secondaries are relaxed with eccentricity close to 0 and synchronous spin state. For readability, spin periods are shortened to five digits and their error bar omitted if significantly smaller than the last written digit.

| Name of primary | Diameter of primary $D_p$ (km) | Size ratio $D_s/D_p$ | Orbital ratio $a_{orb}/D_p$ | Eccentricity $e$ | Primary spin period $P_p$ (hr) | Orbital period $P_{orb}$ (hr) | Secondary spin period $P_s$ (hr) | Age (kyr) |
|---|---|---|---|---|---|---|---|---|
| (3749) Balam | 4.1 ± 0.5 | 0.46 ± 0.05 | 3.1 ± 15% | 0.03-0.08 | 2.8049 | 33.38 ± 0.02 | 33.39 ± 0.02 | $401^{+317}_{-147}$ |
| (3749) Balam | 4.1 ± 0.5 | 0.24 ± ... | 55 ± ... | 0.3-0.8 | 2.8049 | 2600 ± ... | ... | $401^{+317}_{-147}$ |
| (6369) 1983 UC | 3.3 ± 13% | 0.37 ± 0.02 | 3.4 ± 15% | 0 | 2.3971 | 39.8 ± 0.02 | 39.8 ± 0.02 | $671^{+454}_{-349}$ |
| (8306) Shoko | 2.4 ± 13% | ≥ 0.40 (± 0.05) * | 3.3 ± 15% | 0 | 3.3502 | 36.2 ± 0.04 | 36.2 ± 0.04 | $458^{+384}_{-149}$ |
| (9783) Tensho-kan | 5.1 ± 0.6 | 0.24 ± 0.02 | 2.8 ± 15% | 0 | 3.0108 ± 0.0003 | 36.2 ± 0.006 | 36.2† | $671^{+1403}_{-293}$ |
| (10123) Fideoja | 3.2 ± 0.6 | 0.36 ± 0.02 | 4.3 ± 15% | 0 | 2.8662 ± 0.0001 | 56.46 ± 0.02 | 56.46† | ≥ 762 |
| (21436) Chaoyichi | 1.9 ± 0.3 | 0.36 ± 0.02 | 5.5 ± 15% | 0.16-0.22 | 2.8655 ± 0.0002 | 81.19 ± 0.02 | ... | $31^{+109}_{-21}$ |
| (25021) Nischaykumar | 2 ± 0.6 | 0.28 ± 0.03 | 2.4 ± 15% | 0 | 2.5344 ± 0.0012 | 23.4954 ± 0.0004 | 23.5 ± 0.02 | $925^{+1014}_{-416}$ |
| (26416) 1999 XM84 | 3.4 ± 13% | ≥ 0.25 (± 0.06)* | 2.2 ± 15% | 0† | 2.966 ± 0.0001 | 20.785 ± 0.0002 | 20.78 ± 0.01 | $272^{+563}_{-167}$ |
| (26420) 1999 XL103 | 1.2 ± 8% | ≥ 0.34 (± 0.06)* | ~ 3.9 ± 15% | 0 | 3.2 ± 1 | 47.8† ± 0.05 | 47.8† | $252^{+586}_{-107}$ |
| (43008) 1999 UD31 | 1.8 ± 14% | ≥ 0.35 (± 0.05)* | 1.9 ± 15% | 0 | 2.6414 | 16.745 ± 0.005 | 16.7 ± 0.1 | $272^{+460}_{-77}$ |
| (44620) 1999 RS43 | 1.9 ± 13% | 0.39 ± 0.03 | 3.1 ± 15% | 0† | 3.1393 ± 0.0003 | 33.6455 ± 0.0003 | 33.2 ± 0.2 | $742^{+986}_{-549}$ |
| (46829) McMahon | 2.5 ± 13% | 0.4 ± 0.02 | 2 ± 15% | 0† | 2.6236 ± 0.0003 | 16.833 ± 0.002 | 16.833† | $766^{+418}_{-226}$ |
| (80218) 1999 VO123 | 0.9 ± 18% | 0.32 ± 0.02 | 3.1 ± 15% | 0 | 3.1451 ± 0.0002 | 33.1 ± 0.05 | 33.4 ± 0.1 | $143^{+819}_{-44}$ |

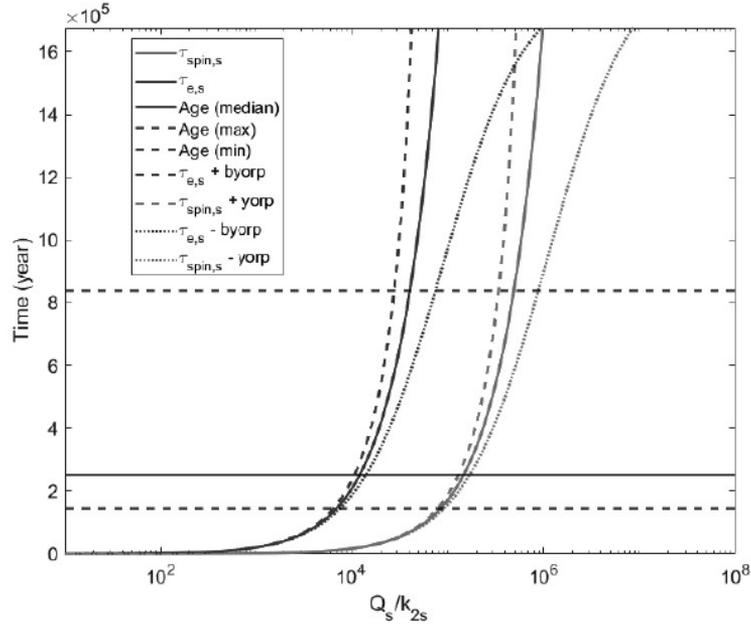

**Figure 1:** Tidal locking timescale $\tau_{\Omega,s}$ and eccentricity damping timescale $\tau_{e,s}$ for the secondary of asteroid (26420) 1999 XL103 based on the values from Pravec et al. (2019). Solid colored lines represent the timescales purely due to tides. The timescales are to be compared to the expected age of the primary on the y-axis from asteroid pairs separation estimation (horizontal black lines) to give an estimate on the tidal dissipation ratio $Q_s/k_{2,s}$ on the x-axis. The impact of the YORP and BYORP effects are shown with dashed colored lines.

### 5.1. Constraints on the tidal dissipation in secondaries of binary systems

For a secondary, three different regimes can be generally identified:

1. if the secondary is tidally locked and on a circular orbit, tidal dissipation was strong enough to make it reach its tidal end states and we get an upper bound on $Q_s/k_{2,s}$;

2. if the secondary is tidally locked but on an eccentric orbit, tidal dissipation was strong enough to alter its spin rate towards synchronous rotation but not enough to damp its eccentricity, thus giving both an upper bound and a lower bound on $Q_s/k_{2,s}$;

3. if the secondary is neither tidally locked nor on a circular orbit, tidal dissipation was too weak to reach either tidal end states, hence providing a lower bound on $Q_s/k_{2,s}$.

Unlike what the extreme cases in Fig. 1 might seem to indicate, the BYORP effect cannot circularize the orbit of the secondary before it is tidally locked because BYORP requires tidal locking to alter the semi-major axis and eccentricity of the system. In the first regime (1), the upper bound on $Q_s/k_{2,s}$ is therefore given by the eccentricity damping timescale $\tau_e$. For regime (2), the upper bound on $Q_s/k_{2,s}$ is provided by the tidal locking timescale $\tau_\Omega$, while the eccentricity timescale $\tau_e$ is used as a lower bound. In the last regime (3), only a lower bound from the tidal locking timescale $\tau_\Omega$ can be derived. Depending on whether an upper bound or lower bound is derived, we use either the maximum age or the minimum age of the asteroid pairs as a constraint on $Q_s/k_{2,s}$. This process is illustrated in Fig. 2. As seen in Table 1, most secondaries reside in the first regime.

The results applied to all secondaries in Table 1 with the exception of (10123) Fideöja are shown in Table 2. The error bars given for the constraint take into account the error measurements from Table 1 on the orbital parameters of the binary systems and other assumptions, but not the error bar on the age of the asteroid pairs. This last error, larger than any other in the table, is considered by taking the upper age value for upper bound constraints, and taking the lower age value for lower bound constraints (as shown in Fig. 2). Details on the error bar calculations in given in section 6.4. For most secondaries, only an upper bound can be derived on their $Q_s/k_{2,s}$. As we will see in section 5.3, these bounds suggest that all secondaries studied are rubble piles, consistent with theories of binary asteroid formation (Walsh et al., 2008a; Jacobson and Scheeres, 2011a).

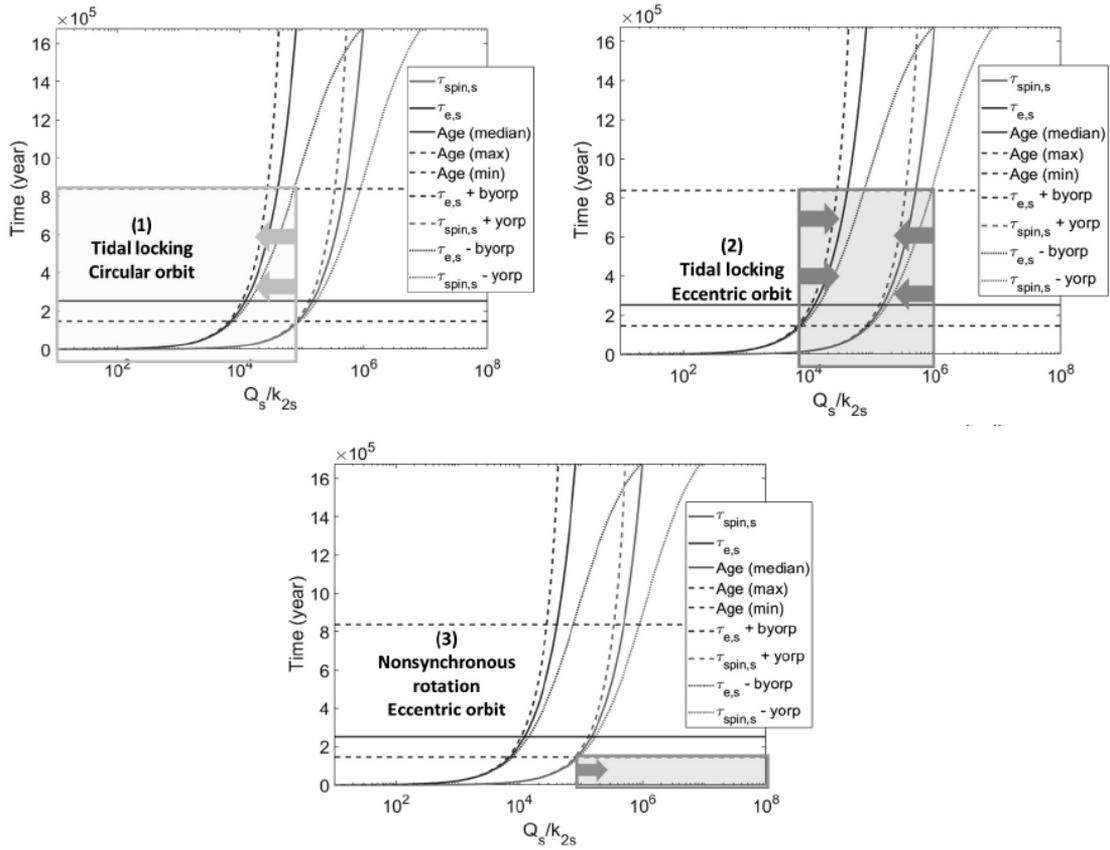

**Figure 2:** Constraints on the tidal dissipation ratio $Q_s/k_{2,s}$ depending on the current state of the binary system, applied to the secondary of asteroid (26420) 1999 XL103. **Top left:** upper bound on $Q_s/k_{2,s}$ for a tidally locked secondary on a circular orbit. **Top right:** upper bound on $Q_s/k_{2,s}$ for a tidally locked secondary, and lower bound on $Q_s/k_{2,s}$ as its orbit is still eccentric. **Bottom:** lower bound on $Q_s/k_{2,s}$ for a non-synchronous secondary on an eccentric orbit. For (26420) 1999 XL103, the relevant figure and process is given in the top left figure corresponding to regime (1).

Although in most cases only an upper or lower bound can be established, one exception is the first bound secondary of (3749) Balam which is synchronous but on an eccentric orbit. In this case it is possible to get an actual measurement of $Q_s/k_{2,s}$ from the upper bound due to tidal locking and the lower bound from eccentricity damping. We find that the tidal dissipation ratio of the first secondary of asteroid (3749) Balam is between $2.8 \times 10^4$ and $1.5 \times 10^6$. For the second secondary of (3749) Balam and the secondary of (21436) Chaoyichi, only a lower bound can be put on their tidal dissipation ratio. Because these secondaries are very young, tidal dissipation has not occurred for long enough to either tidally lock them or circularize their orbit.

### 5.2. Constraints on the tidal dissipation in primaries of binary systems

As we only consider the impact of tidal dissipation in the primaries regarding their change in spin rate, only two regimes are considered:

1. if the primary is tidally locked, tidal dissipation was strong enough to make it reach its tidal end states and we get an upper bound on $Q_p/k_{2,p}$;

Table 2

Constraints on the tidal dissipation of secondaries in binary systems inside asteroid pairs. Regime (1) means that the secondary is tidally locked and on a circular orbit. Regime (2) is for secondaries that are tidally locked but on an eccentric orbit. Regime (3) corresponds to secondaries that are neither in synchronous rotation nor on a circular orbit. The reason for the large error bar of the first bound secondary of (3749) Balam is discussed in section 6.1. (80218) 1999 VO123 has no defined upper error bar, as for extreme values of its orbital parameters a maximal, circularizing BYORP effect alone would be strong enough to damp the eccentricity of the system in its maximal age, leading to an undetermined $Q_s/k_{2s}$. Results from TM11 (Taylor and Margot, 2011) considering only tidal end states and no thermal torques are given for reference.

| Primary asteroid | $R_s$ (m) | Tidal state regime | $Q_s/k_{2s}$ constraint | $Q_s/k_{2s}$ from TM11 |
|---|---|---|---|---|
| (3749) Balam | 943 | (2) | $2.7 \times 10^4 \,^{+4.6 \times 10^4}_{-2.5 \times 10^4} \leq Q_s/k_{2s} \leq 1.4 \times 10^6 \,^{+3.9 \times 10^6}_{-0.48 \times 10^6}$ | $\geq 2.5 \times 10^5$ |
| (3749) Balam | 492 | (3) | $\geq 1.2 \times 10^{-2} \,^{+6.6 \times 10^{-2}}_{-4.5 \times 10^{-2}}$ | $\geq 2.6 \times 10^{-4}$ |
| (6369) 1983 UC | 611 | (1) | $\leq 1.6 \times 10^5 \,^{+2.9 \times 10^5}_{-0.87 \times 10^5}$ | $\leq 5.9 \times 10^5$ |
| (8306) Shoko | 480 | (1) | $\leq 1.9 \times 10^5 \,^{+5.3 \times 10^5}_{-0.98 \times 10^5}$ | $\leq 6.9 \times 10^5$ |
| (9783) Tensho-kan | 612 | (1) | $\leq 4.5 \times 10^5 \,^{+8.3 \times 10^5}_{-2.6 \times 10^5}$ | $\leq 9.3 \times 10^5$ |
| (21436) Chaoyichi | 342 | (3) | $\geq 6.5 \times 10^2 \,^{+4.1 \times 10^2}_{-2.3 \times 10^2}$ | $\geq 1.5 \times 10^2$ |
| (25021) Nischaykumar | 280 | (1) | $\leq 2.2 \times 10^6 \,^{+7.6 \times 10^6}_{-1.4 \times 10^6}$ | $\leq 3.9 \times 10^6$ |
| (26416) 1999 XM84 | 425 | (1) | $\leq 9.3 \times 10^5 \,^{+40 \times 10^5}_{-4.7 \times 10^5}$ | $\leq 2.1 \times 10^6$ |
| (26420) 1999 XL103 | 204 | (1) | $\leq 7.5 \times 10^4 \,^{+22 \times 10^4}_{-4.0 \times 10^4}$ | $\leq 1.4 \times 10^5$ |
| (43008) 1999 UD31 | 315 | (1) | $\leq 4.4 \times 10^6 \,^{+13 \times 10^6}_{-2.3 \times 10^6}$ | $\leq 1.4 \times 10^7$ |
| (44620) 1999 RS43 | 371 | (1) | $\leq 7.1 \times 10^5 \,^{+14 \times 10^5}_{-5.0 \times 10^5}$ | $\leq 2.0 \times 10^6$ |
| (46829) McMahon | 500 | (1) | $\leq 6.9 \times 10^6 \,^{+10 \times 10^6}_{-3.8 \times 10^6}$ | $\leq 2.6 \times 10^7$ |
| (80218) 1999 VO123 | 144 | (1) | $\leq 1.3 \times 10^6 \,^{+\infty}_{-1.0 \times 10^6}$ | $\leq 5.7 \times 10^5$ |

2. if the primary is not in synchronous rotation, tidal dissipation was too weak to reach either tidal end state, hence providing a lower bound on $Q_p/k_{2,p}$.

Because of the mass and size difference between the primaries and their respective secondaries, the secondary gets tidally locked much faster than the primary (Eq. 5 and Eq. 6). This is confirmed by the observational data in Table 1, where while most secondaries are tidally locked (with the exception of (3749) Balam's second secondary and (21436) Chaoyichi's secondary), none of the primaries are. All primaries are therefore in the first regime, and only a lower bound on $Q_p/k_{2,p}$ can be derived: the values are given in Table 3. As we will see in section 5.3, while the values of the lower bounds are also compatible with monoliths, the trend of these lower bounds with increasing radius suggests that all primaries are also rubble piles.

## 5.3. Comparison with previous studies
### 5.3.1. Goldreich and Sari 2009

Goldreich and Sari (2009) studied the difference in rigidity between a monolithic body and a rubble pile of the same composition, assuming tidal forces weaker than self gravity and that stress concentrations and plastic yielding are important due to voids rather than cracks. In the case of a monolithic body, classical tidal dissipation for an uniform elastic body gives the following expression of $k_2$ (Munk and MacDonald, 1960):

$$(k_2)_{monolithic} = \frac{3}{2}\left(1 + \frac{19\mu}{\rho g R}\right)^{-1} \tag{10}$$

where $\mu$ and $g$ are the body's rigidity and surface gravity, respectively. In the case of small bodies, $\mu \gg \rho g R$ hence $k_2$ scales as $1/\mu$, with $\mu$ typically being in the order of magnitude of a GPa or larger for rocky bodies. For monoliths, $Q$ is typically assumed to be 100 (Goldreich and Soter 1966) hence, assuming $\mu = 10^{10}$ Pa:

$$\left(\frac{Q}{k_2}\right)_{monolithic} \approx 10^{16} R^{-2} \approx 10^{10}\left(\frac{R}{1\,km}\right)^{-2} \tag{11}$$

**Table 3**
Constraints on the tidal dissipation of primaries in binary systems inside asteroid pairs. Regime (1) means that the primary is tidally locked. Regime (2) is for primaries that are not in synchronous rotation, which is all of them in this work. As this table considers primaries only, (3749) Balam is not present twice. Results from JS11 (Jacobson and Scheeres, 2011b) assume B = $10^{-2}$. Results from TM11 (Taylor and Margot, 2011) considering only tidal end states and no thermal torques are given for reference.

| Primary asteroid | $R_p$ (m) | Tidal state regime | $Q_p/k_{2p}$ constraint | $Q_p/k_{2p}$ from JS11 | $Q_p/k_{2p}$ from TM11 |
|---|---|---|---|---|---|
| (3749) Balam | 2050 | (2) | $\geq 1.7 \times 10^3 \,^{+3.4\times10^3}_{-1.1\times10^3}$ | $4.1793 \times 10^5$ | $\geq 2.5 \times 10^5$ |
| (6369) 1983 UC | 1650 | (2) | $\geq 3.9 \times 10^2 \,^{+4.8\times10^2}_{-2.1\times10^2}$ | $6.2093 \times 10^4$ | $\geq 1.2 \times 10^5$ |
| (8306) Shoko | 1200 | (2) | $\geq 1.1 \times 10^3 \,^{+5.3\times10^3}_{-0.38\times10^3}$ | $5.2098 \times 10^4$ | $\geq 1.8 \times 10^5$ |
| (9783) Tensho-kan | 2550 | (2) | $\geq 1.4 \times 10^2 \,^{+2.4\times10^2}_{-0.84\times10^2}$ | $1.3916 \times 10^5$ | $\geq 1.4 \times 10^5$ |
| (21436) Chaoyichi | 950 | (2) | $\geq 0.70 \,^{+0.9}_{-0.37}$ | $5.83 \times 10^2$ | $\geq 1.5 \times 10^2$ |
| (25021) Nischaykumar | 1000 | (2) | $\geq 1.0 \times 10^3 \,^{+2.0\times10^3}_{-0.66\times10^3}$ | $8.8001 \times 10^4$ | $\geq 8.0 \times 10^5$ |
| (26416) 1999 XM84 | 1700 | (2) | $\geq 2.1 \times 10^2 \,^{+33\times10^2}_{-0.72\times10^2}$ | $3.0428 \times 10^5$ | $\geq 2.1 \times 10^5$ |
| (26420) 1999 XL103 | 600 | (2) | $\geq 6.5 \times 10^1 \,^{+5.6\times10^1}_{-2.2\times10^1}$ | $2.07 \times 10^2$ | $\geq 1.7 \times 10^5$ |
| (43008) 1999 UD31 | 900 | (2) | $\geq 6.1 \times 10^3 \,^{+37\times10^3}_{-2.1\times10^3}$ | $9.0581 \times 10^5$ | $\geq 2.8 \times 10^6$ |
| (44620) 1999 RS43 | 950 | (2) | $\geq 7.7 \times 10^2 \,^{+12\times10^2}_{-4.5\times10^2}$ | $4.3544 \times 10^4$ | $\geq 1.6 \times 10^5$ |
| (46829) McMahon | 1250 | (2) | $\geq 3.1 \times 10^4 \,^{+3.6\times10^4}_{-1.6\times10^4}$ | $2.1765 \times 10^5$ | $\geq 8.3 \times 10^6$ |
| (80218) 1999 VO123 | 450 | (2) | $\geq 1.3 \times 10^2 \,^{+17\times10^2}_{-0.73\times10^2}$ | $4.667 \times 10^3$ | $\geq 4.4 \times 10^4$ |

where $R$ is in SI units. In the case of a rubble pile asteroid, Goldreich and Sari (2009) calculated that its tidal response is given by:

$$(k_2)_{rubble} \approx \rho R \left(\frac{G\epsilon_Y}{\mu}\right)^{1/2} \quad (12)$$

where $\epsilon_y$ is the yield strain with a nominal value of $\epsilon_y = 10^{-2}$. They however did not derive any scaling law for the tidal dissipation factor $Q$ and used the same assumed $Q \approx 10^2$ from monolithic bodies. Assuming typical values for the bulk density of asteroid $\rho = 2000$ kg.m$^{-3}$ and $\mu = 10^{10}$ Pa, their prediction on the tidal dissipation ratio $Q/k_2$ for rubble pile becomes:

$$\left(\frac{Q}{k_2}\right)_{rubble} \approx 6 \times 10^9 R^{-1} \approx 6 \times 10^6 \left(\frac{R}{1\ km}\right)^{-1} \quad (13)$$

Eq. 11 and Eq. 13 are compared to our results from Table 2 and Table 3 on Fig. 3. The derived values from Goldreich and Sari (2009) for tidal $Q/k_2$ for monolithic bodies (eq. 11) are much higher than the upper bounds we put on $Q_s/k_{2s}$, arguing that no secondary is a monolith and suggesting that they are rather rubble piles. However, the $Q/k_2$ values for rubble pile asteroids using the theory of Goldreich and Sari (2009) (eq. 13) are also still higher than the upper bounds we get on $Q_s/k_{2s}$. We will see in the following section that this result suggests the theory of Goldreich and Sari (2009) is incomplete and must take into account other dependencies, such as a non-constant $Q$ or not being governed by continuum mechanics as assumed there. For primaries, the lower bounds on $Q_p/k_{2p}$ are consistent with both rubble piles and monoliths.

### 5.3.2. Jacobson and Scheeres 2011

When Goldreich and Sari (2009) studied the impact of YORP and BYORP on the orbital elements of a binary system, they noted that a possible equilibrium could exist for rubble pile binaries with synchronous secondaries between the semi-major axis evolution due to tides and due to the BYORP effect. This equilibrium has been studied in detail in Jacobson and Scheeres (2011b) and used to study the tidal dissipation ratio of binary asteroid systems likely to be in such an equilibrium.

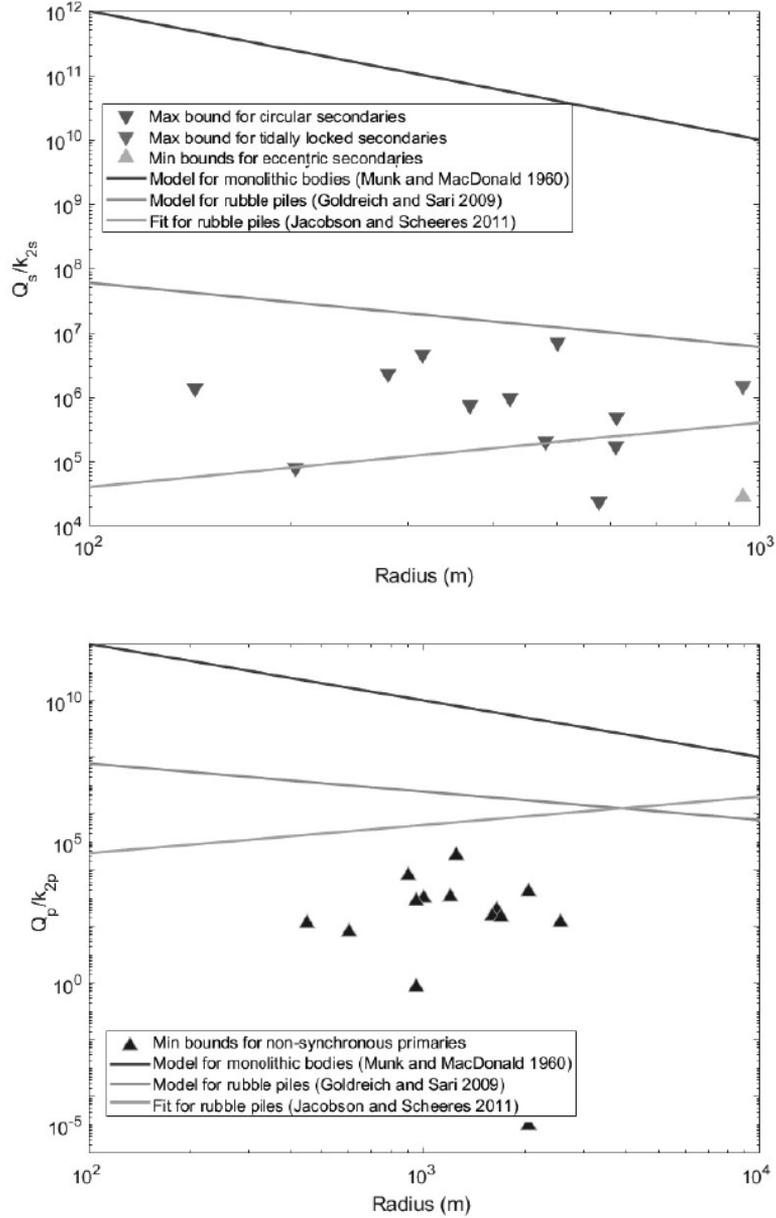

**Figure 3:** Comparison between our constraints on $Q/k_2$ for binary systems in asteroid pairs and the $Q/k_2$ estimations from Munk and MacDonald (1960) for monoliths, and from Goldreich and Sari (2009) and Jacobson and Scheeres (2011b) for rubble pile. **Top figure** is for secondaries in Table 2. **Bottom figure** is for primaries in Table 3. Results are consistent with the fit from Jacobson and Scheeres (2011b).

For binaries in a tides/BYORP equilibrium, the ratio $BQ_p/k_{2p}$ is given by (Jacobson and Scheeres, 2011b):

$$\left(\frac{BQ_p}{k_{2p}}\right)_{equilibrium} = \frac{2\pi \omega_d^2 \rho R_p^2 q^{4/3}}{H_\odot} \left(\frac{R_p}{a}\right)^7 \tag{14}$$

*Tidal dissipation of binaries in asteroid pairs*

with the same notation used in section 3.2 and $H_{\ast} = F_S/a_S^2 \sqrt{1-e_S^2}$. While their nominal value was $B = 10^{-3}$, a more likely value is the maximum value used in their figure $B = 10^{-2}$ as pointed out by Scheirich et al. (2015). As they assumed $B$ to be constant and $Q \approx 10^2$, their fit with observational data can be rewritten:

$$\left(\frac{Q_p}{k_{2p}}\right)_{equilibrium} \approx 4 \times 10^2 R_p \approx 4 \times 10^5 \left(\frac{R_p}{1\ km}\right) \tag{15}$$

Eq. 15 is compared to our results from Table 2 and Table 3 on Fig. 3. For the primaries, all lower bounds we derive are compatible with the fit from Jacobson and Scheeres (2011b), as their expected $Q_p/k_{2p}$ is greater than our lower bounds, but the trend of our lower bounds seems to be in good agreement with them. This suggests that $Q_p/k_{2p}$ for rubble piles would indeed tend to be increasing with radius, unlike the theory from Goldreich and Sari (2009). While the Jacobson and Scheeres (2011b) theory was mainly on primaries, this fit is interestingly compatible with almost all the bounds we get for the secondaries in binary asteroids too. One must note that while a weaker BYORP effect would improve the quality of fit for the secondaries, it would on the other hand worsen the fit for the primaries; at the same time, this fit is only true if the binary system is in the BYORP and tidal equilibrium described in Jacobson and Scheeres (2011b). All in all, our results compared with Jacobson and Scheeres (2011b) seem to indicate that all bodies studied in this work are indeed rubble piles, and that $Q/k_2$ increases with increasing radius for rubble pile asteroids.

### 5.3.3. Nimmo and Matsuyama 2019

Nimmo and Matsuyama (2019) assumed that for a rubble pile asteroid, tidal dissipation occurs only in a surface regolith layer of unknown thickness $t$ and is due to sliding friction during relative displacement between neighboring elements. Comparing the total frictional dissipation rate to the conventional tidal dissipation in a non-synchronous primary, they derived an effective $Q$ for rubble pile asteroids:

$$\left(Q_{\text{eff}}\right)_{rubble} \approx \frac{qn^2}{G\rho}\frac{1}{Nf}\left(\frac{R_p}{t}\right)^2 \tag{16}$$

where $q$ is the mass ratio $m_s/m_p$ and $N$ is the half the number of faces per element experiencing friction in the dissipative surface layer of thickness $t$ and $f$ is the friction coefficient during sliding. For silicates, $f$ is around 0.6 while $N$ is roughly 3 for cubic elements; hence, $Nf$ can be considered of order unity. Assuming $\rho = 2000$ kg.m$^{-3}$, this expression becomes:

$$\left(Q_{\text{eff}}\right)_{rubble} \approx 1 \times 10^7\ qn^2 \left(\frac{R_p}{t}\right)^2 \tag{17}$$

Combining Eq. 17 with Eq. 12 based on Goldreich and Sari (2009) gives the following prediction for the tidal dissipation ratio:

$$\left(\frac{Q_{\text{eff}}}{k_2}\right)_{rubble} \approx 1 \times 10^{15}\ qn^2 R_p \left(\frac{1}{t}\right)^2 \tag{18}$$

Eq. 18 is compared to our results from Table 2 and Table 3 in Fig. 4. In the case of primaries, all derived $Q_{\text{eff},p}/k_{2p}$ are compatible with our lower bounds, suggesting that all primaries are rubble piles and can have a surface regolith layer as thick as 100 m. On the other hand, the dissipation $Q_{\text{eff},s}/k_{2s}$ associated with a surface layer of only 3 m is not enough to match the tidal dissipation $Q_s/k_{2s}$ in almost all studied secondaries. This would imply that secondaries are likely to have a surface dissipative layer of a couple tens of meters at least. Of note is the case of the first bound secondary of (3749) Balam, where having a 100 m-thick surface layer would make tidal dissipation almost stronger than the observations, suggesting that this secondary has a dissipative surface layer between 30 m and 100 m thick. The general good agreement with Nimmo and Matsuyama (2019) suggests that $Q/k_2$ does scale with increasing radius for rubble piles. These results are however assuming the tidal dissipation theory of Nimmo and Matsuyama (2019) can also be applied to synchronous secondaries.

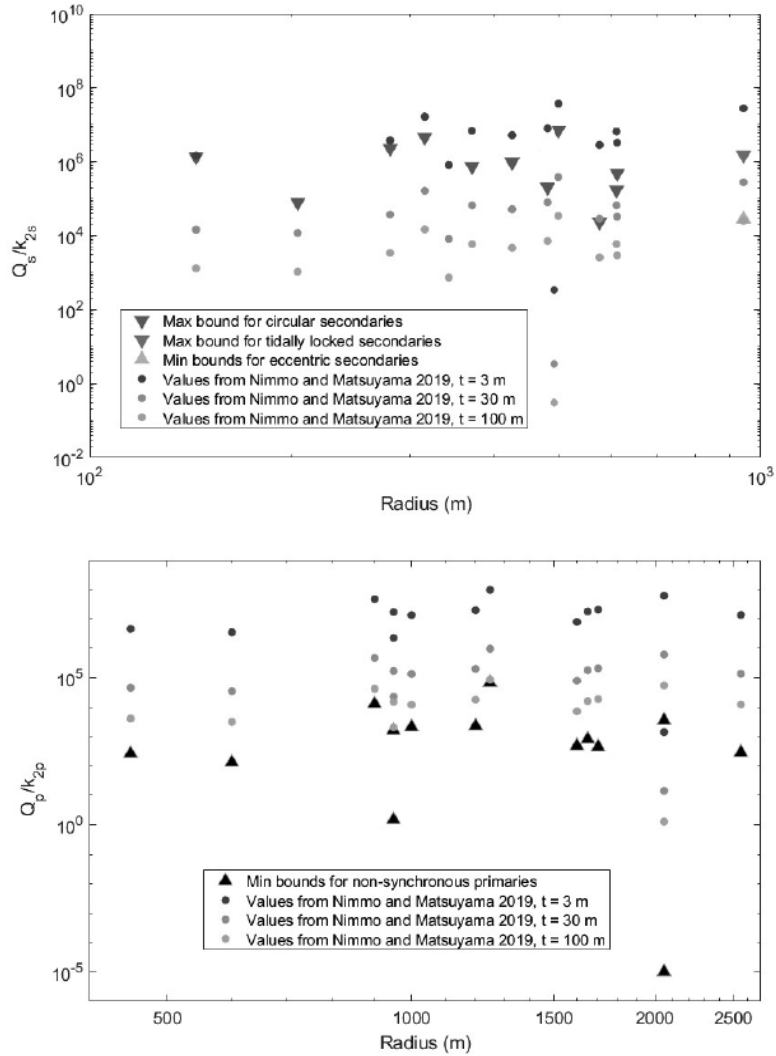

**Figure 4:** Comparison between our constraints on $Q/k_2$ for binary systems in asteroid pairs and the theoretical $Q_{\text{eff}}/k_2$ values from Nimmo and Matsuyama (2019). **Top figure** is for secondaries from Table 2. **Bottom figure** is for primaries from Table 3. Different values of regolith thickness for $Q_{\text{eff}}/k_2$ have been used: 3 m, 30 m, or 100 m. While primaries can fit any thickness with little trend compared to radius, the upper bounds on secondaries favor an intermediate thickness around 30 m or more. No conclusive trend on the regolith thickness with respect to radius can be inferred from our data.

## 6. Discussion

When estimating upper bounds and lower bounds on the tidal dissipation inside primaries and secondaries of binary or multiple asteroid systems, several assumptions were made. By using the tidal timescales in section 2, we assume in this work that the binary did not form in its tidal end state, as the age of the asteroid pair could not be used to constrain the parameters of the remaining binary in that case. As this work is mainly about order-of-magnitude rather than accurate individual predictions on binary asteroids, dynamical modeling of the secondary formation to accurately determine its initial spin state and eccentricity goes beyond the scope of this work and we use arbitrary initial spin and eccentricity values based on actual observations given in Table 1.

In section 2, we assumed only binaries (neglecting the effect of other secondaries in the system) that evolved on an orbit with no inclination $i$ and either no eccentricity $e$ or neglecting the terms of order 2 in eccentricity, and that the angular momentum was conserved. We also restricted our study to degree-2 tides and used a constant phase lag model with $Q$ independent of amplitude and frequency. Finally, the impact of tides inside the primary was omitted when calculating the change in eccentricity due to tides. We wish to discuss these assumptions in the following section.

The tidal evolution of the spins and the eccentricity of the binary system in section 2 assumes conservation of angular momentum. This means that we are neglecting any large perturbations from outside the system, such as impacts or close encounters with other celestial bodies. Relaxing this assumption would require accurate estimation on the evolution of the angular momentum of the binary with respect to time. This notably means having ways to image the surface of these bodies looking for large crater impacts, and orbital evolution tools to compare the orbital history of the studied bodies with known objects. The YORP and BYORP effects also break angular momentum conservation, and therefore a better estimation of the spin and eccentricity evolution timescales would need dynamical modeling incorporating both tides and thermal effects of the binary systems. Such studies go beyond the scope of this work and therefore were neglected.

While commonly assumed to be constant, tidal dissipation $Q$ is in fact most likely a function of frequency (e.g., Efroimsky and Lainey, 2007; Boué and Efroimsky, 2019). However, as illustrated with the comparison of the previous works from Goldreich and Sari (2009) or Nimmo and Matsuyama (2019), tidal dissipation in rubble piles is to this day still ill-constrained. When also considering the large spread of possible ages of the asteroid pairs used in this work and the error from the measurements in Table 1 (see also section 6.4), we concluded that it was reasonable to mesh the unknown frequency dependence of $Q$ in these error bars given how large they already were.

### 6.1. Impact of primary tides on eccentricity

While tides in the secondary act to damp the eccentricity of its orbit, tides in the primary instead increase the eccentricity of the secondary's orbit. From tidal theory, the rate of change in eccentricity due to tides in the primary is Murray and Dermott (2000); Jacobson and Scheeres (2011b):

$$\dot{e}_{\text{primary}} = \frac{57}{8} \frac{k_{2p}}{Q_p} \left(\frac{R_s}{a}\right)^5 \left(\frac{m_p}{m_s}\right)^{\frac{2}{3}} ne \qquad (19)$$

Because tides in the primary act against tides in the secondary with regards to eccentricity, the total change in eccentricity $\dot{e}$ will in fact be smaller. This means that the real eccentricity damping timescale $\tau_e$ will be greater than the one derived in Eq. 7: the stronger tidal dissipation is in the primary, the greater the difference in timescales. If circularization has occurred, this means that we will have in fact underestimated tidal dissipation in the secondary, thus overestimated $Q_s/k_{2s}$. However, in most cases the eccentricity damping timescale is used to put upper bounds on $Q_s/k_{2s}$: therefore the worst case is to overestimate it. As such, the most conservative upper bound on $Q_s/k_{2s}$ is when the primary tides are neglected, which is what was done in the previous sections for most secondaries. A single exception here is the case of the first bound secondary of asteroid (3749) Balam, where the eccentricity damping timescale is used as a lower bound on $Q_s/k_{2s}$. For this case only, primary tides were considered and displayed in the error bar on the lower bound for this secondary in Table 2. The impact of primary tides on the eccentricity timescale is also shown in Fig. 5.

In the case of asteroid (3749) Balam, we also considered the two secondaries separately. While interactions between them could alter their orbital evolution, we chose to neglect those perturbations because of the large separation between them ($a_{orb}/D_p$=3.1 and 55, respectively).

### 6.2. Higher-degree tides for close binaries

Taylor and Margot (2011) studied the impact of tides in binary asteroid systems to determine their tidal end states and how to use it to constrain their physical properties. Assuming formation of a binary asteroid system is likely to come from a single parent body being disrupted either by thermal torques (e.g., Bottke, 2002; Scheeres, 2007), close planetary encounters (e.g., Richardson et al., 1998), or large impacts (Durda et al., 2004), they constrain the tidal dissipation in binaries based on their age since separation and current value of their semi-major axis.

Using the age of the asteroid pair for $\Delta t$, the comparison between the results from Taylor and Margot (2011) are given in the last column of Table 2 and Table 3. Since our tidal approaches are similar but we also took into account the effect of YORP and BYORP on orbital evolution, they may underestimate tidal dissipation in the primaries resulting

Tidal dissipation of binaries in asteroid pairs in their lower bounds being larger than ours. The comparison of our methods on secondaries is more interesting: at the first glance, the upper bounds on $Q_s/k_{2s}$ seem to be lower than ours, which would be expected due to the omission of tidal torques, and would suggest that the separation distance used at the formation of a binary system from Walsh et al. (2008b) is a good approximation. However, a few upper bounds derived from Taylor and Margot (2011) are almost equal or greater than ours, which contradicts our consideration of the radiative torques. This is probably due to the factor that they assume $\mu_p Q_p = \mu_s Q_s$, which is likely wrong when considering the size difference between primaries and secondaries; Eq. 15 and Eq. 18 notably show that $\mu Q$ is expected to depend on radius. This is further supported by the fact that this anomaly is the greatest for the smallest studied secondary.

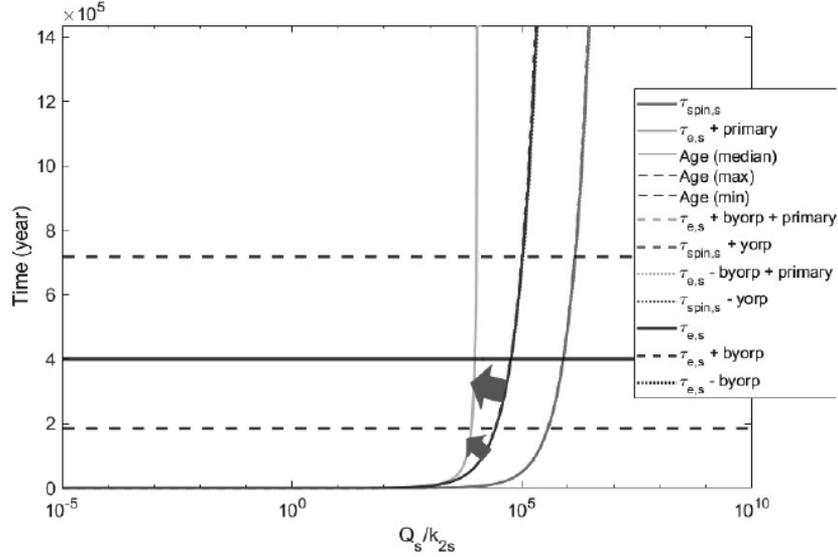

**Figure 5:** Impact of the primary tides on the eccentricity damping timescale for the first bound secondary of asteroid (3749) Balam. Blue lines indicate the original eccentricity damping timescale with no primary tides, while the yellow lines include eccentricity changes due to primary tides. This causes the eccentricity damping timescale to increase, hence decreasing our estimations on $Q_s/k_{2s}$. As tidal dissipation in the secondary becomes weaker (i.e., $Q_s/k_{2s}$ increases), tides in the primary dominate the eccentricity evolution and eccentricity is no longer damped, thus $\tau_e$ becoming vertical. BYORP and YORP effects are also considered but are too small to be visible here. (For interpretation of the reference to color in this figure legend, the reader is referred to the web version of this article.)

While our approach only consider degree-2 tides, Taylor and Margot (2010) also studied the impact of tides up to degree 6 for such binary systems. They showed that the relative contribution of all tides up to degree 6 is at most 30% greater than degree-2 tides when the separation distance $\left(R_p/a\right) = 0.5$, and falls below 1% when $\left(R_p/a\right) = 0.1$. In Table 1, the closest separation distance in $\left(R_p/a\right) = 0.26$ for the binary system of asteroid (43008) 1999 UD31, meaning according to Fig. 5 in Taylor and Margot (2010) that tides up degree 6 contribute to less than 8% more than degree-2 tides. This is much lower than the uncertainties on the age of asteroid pairs and notably lower than the measurement error on $a/D_p$ from Table 1, justifying our restriction to degree-2 tides only.

### 6.3. Higher-order eccentricity terms and orbital elements evolution

The most likely origin of asteroid pairs and binary asteroid systems in rotational fission of a primary due to YORP spin-up (Scheeres, 2007; Pravec et al., 2019). This makes it likely that the initial inclination of the binary system is low, with the secondary being ejected from the equator where stresses are the greatest (e.g., Holsapple, 2007, 2008; Hirabayashi and Scheeres, 2015). However, it also makes likely for it to have non-negligible eccentricity at its formation. To evaluate the impact of high initial eccentricity, we use the works of Boué and Efroimsky (2019) where they reworked the tidal evolution of Kepler elements in a more general case which includes binary asteroid systems. For simplification, we will however assume that both the Love number $k_2$ and the tidal dissipation factor $Q$ are independent of frequency; this is reasonable as this study is only an order-of-magnitude study. Details of the derivation of the change in eccentricity for degree-2 tides up to order 3 in eccentricity is given in Appendix A.1, resulting in the following equation:

$$\dot{e} = (\dot{e})_{primary} + (\dot{e})_{secondary}$$
$$= -ne\frac{m_s}{m_p}\left(\frac{R_p}{a}\right)^5 \frac{k_{2p}}{Q_p}\left(-\frac{57}{8} - \frac{1509e^2}{32}\right) \quad (20)$$
$$- ne\frac{m_p}{m_s}\left(\frac{R_s}{a}\right)^5 \frac{k_{2s}}{Q_s}\left(\frac{21}{2} + \frac{903e^2}{4}\right)$$

The eccentricity term of order 3 for primary tides is greater than the order 1 term for eccentricity $e$ larger than 0.3887, while the term of order 3 for secondary tides requires an eccentricity $e$ greater than 0.4313 to dominate. As even the very young binary asteroid system (21436) Chaoyichi only has an eccentricity smaller than 0.22, those terms are unlikely to matter for a significant duration in our considered binary systems.

Spin evolution is derived from the semi-major axis evolution from Boué and Efroimsky (2019), also detailed in Appendix A.2. The equations governing spin evolution of the primary and the secondary are, in the worst case for initial spin conditions:

$$\dot{\Omega}_p = -\frac{1}{2}\frac{k_{2p}}{\alpha_p Q_p}\frac{m_s^2}{m_p(m_p + m_s)}\left(\frac{R_p}{a}\right)^3 n^2\left(3 + \frac{153}{4}e^2\right) \quad (21)$$

$$\dot{\Omega}_s = -\frac{1}{2}\frac{k_{2s}}{\alpha_s Q_s}\frac{m_p^2}{m_s(m_s + m_p)}\left(\frac{R_s}{a}\right)^3 n^2\left(-3 - \frac{171}{4}e^2\right) \quad (22)$$

corresponding to $\frac{3}{2}n \leq \Omega_p$ and $\Omega_s \leq n/2$. The eccentricity terms of order 2 are stronger than the order-0 terms for $e = 0.2801$ for $\Omega_p$, and for $e = 0.2649$ for $\Omega_s$. The eccentricity values are compatible with the upper bound of the eccentricity of (21436) Chaoyichi and lower bound of the eccentricity of the second secondary of (3749) Balam, meaning that they may need to be considered to get an accurate estimation of the real tidal locking timescale. In both case, the high eccentricity terms contribute to increase $\dot{\Omega}$ i.e., reducing $\tau_\Omega$, hence leading to larger possible $Q/k_2$. Therefore, this only need to be considered when dealing with $\tau_\Omega$ being used as an upper bound, which is only the case for the first bound of asteroid (3749) Balam where an additional factor 2 were added to its error bar.

### 6.4. Error bars on tidal dissipation ratio bounds

To derive the error bars on our bounds in Table 2 and Table 3, we take into account the following errors:

1. error measurements from Table 1 on $R_p$, $a/R_p$, $n$ from $P_{orb}$, $R_s/R_p$. Error on $R_s$, $a$, and mass ratios terms are derived from the previously mentioned parameters, meaning we are assuming that primaries and secondaries have similar shapes and densities;
2. for the first bound secondary of asteroid (3749) Balam, the impact of primary tides on eccentricity (see section 6.1);
3. an additional 8% error bar on upper bounds due to higher-degree tides being neglected (see section 6.2);
4. a factor of 2 on the bounds derived from the tidal locking timescale due to high eccentricity terms, impacting the upper bound for the estimate of $Q_{2s}/k_s$ for the first bound secondary of asteroid (3749) Balam (see section 6.3).

Uncertainties on the age of the binary system alter directly the values in Table 2 and Table 3 as described in section 5.1. Because the initial spin rate of the primary and the secondary are not known, no error is taken on $\Omega_{init}$ from $P_p$; the same is done for $\alpha_p$ as the actual shape of the bodies is not accurately known. The error measurements from Table 1 are considered by using their lower and higher values instead of their mean values and recalculating the tidal and thermal timescales with these values. Item 3 and 4 are then added on top of the newly calculated bounds to give the final error bars displayed in Table 2 and Table 3.

While librations due to the eccentricity of the secondary's orbit are considered for tidal timescales, they can also alter the BYORP effect. Ćuk and Nesvorný (2010) notably showed that BYORP can cause librations to be chaotic, and these librations can in turn halt BYORP. While these complex dynamics are not modeled here, the possibility of having librations turn off BYORP would mean a lower BYORP effect than expected over the life of the binary. As we consider a wide range of BYORP coefficient between $+10^{-2}$ and $-10^{-2}$, this effect is already implicitly accounted for in our analysis.

We also considered eccentricity damping in section 2.3 until the orbit is fully circular, i.e., e=0, which is the classical tidal end state. However, because of non-Keplerian dynamics in binary asteroids and other gravitational influences, the actual equilibrium eccentricity may be nonzero and therefore reached earlier, meaning a smaller $Q/k_2$. As we are in all relevant cases using the eccentricity timescale as an upper bound, possible nonzero equilibrium eccentricities are valid with our derived upper bounds on $Q_s/k_{2s}$.

## 7. Conclusion

Tidal theory and radiative torques from BYORP and YORP inside binary asteroids systems can be used to probe the interiors of asteroids inside binary systems. Using the age constraints of asteroid pairs, it is possible to estimate the maximal age of binary systems existing inside these pairs. From these age estimates, we derive bounds on the tidal dissipation ratio $Q/k_2$ of primaries and secondaries of these binary systems. The results for primaries are not well-constrained, being compatible with both monolithic rocks and rubble piles. However, we show that all secondaries are consistent with rubble piles based on the tidal dissipation theories in rubble piles from Jacobson and Scheeres (2011b) and Nimmo and Matsuyama (2019). This result is consistent with existing theories of binary asteroid formation (Walsh et al., 2008a; Jacobson and Scheeres, 2011a). Our upper bounds on tidal dissipation inside these secondaries suggest that $Q/k_2$ increases with radius, consistent with the results from Jacobson and Scheeres (2011b) on primaries in BYORP equilibrium, and imply that they are covered with regolith layers at least a few tens of meters thick.

A notable case is the first bound secondary of asteroid (3749) Balam. Due to tidally locked but on an eccentric orbit, we estimate its $Q/k_2$ ratio to be between $2.7 \times 10^4$ and $1.4 \times 10^6$, constraining it as a rubble pile with a regolith layer between 30m and 100m thick.

Our work is currently limited to binary asteroids inside asteroids pairs, as the age of the binary system must be constrained to infer tidal dissipation from their orbital configuration. More accurate measurements of the orbital parameters of the binary systems, dynamical modeling of their formation and evolution could also improve these estimations, with high-order eccentricity terms notably impacting the tidal locking timescales which is used to derive the upper bounds on the secondaries $Q/k_2$. Future missions towards asteroid binaries, such as Lucy (Levison et al., 2021) with the (3548) Eurybates and the (617) Patroclus–Menoetius binary systems and Hera (Michel et al., 2018) with the (65803) Didymos system will probe the interior structure of these asteroids, allowing to better understand the relation between tidal dissipation and regolith thickness, as well as tying binary system age estimations with tides and thermal torques.

## Acknowledgements

L. Pou was supported by appointments to the NASA Postdoctoral Program at the NASA Jet Propulsion Laboratory, California Institute of Technology, administered by Oak Ridge Associated Universities under contract with NASA (80HTQ21CA005).

## CRediT authorship contribution statement

**Laurent Pou:** Conceptualization, Methodology, Software, Writing. **Francis Nimmo:** Conceptualization, Methodology, Writing.

# A. Higher-order eccentricity terms derivation

## A.1. Derivation of the eccentricity evolution

Boué and Efroimsky (2019) derived the tidal evolution of the Keplerian elements for general binary systems where the mass of the secondary is comparable to the mass of the primary. From their Equation (155), the change in eccentricity $e$ can be written to order 5 in eccentricity and order 2 in inclination as:

$$\begin{aligned}
\dot{e} &= (\dot{e})_{primary} + (\dot{e})_{secondary} \\
&= -ne\frac{m_s}{m_p}\left(\frac{R_p}{a}\right)^5 \Bigg(-(1-\frac{e^2}{4})\frac{3}{16}\, K_{2p}(n-2\Omega_p) \\
&\qquad -\frac{3}{4}(1-\frac{21e^2}{4})\, K_{2p}(2n-2\Omega_p) \\
&\qquad +\frac{147}{16}(1-\frac{179e^2}{28}\, K_{2p}(3n-2\Omega_p) \\
&\qquad +\frac{867}{8}e^2\, K_{2p}(4n-2\Omega_p) \\
&\qquad +\frac{9}{8}(1+\frac{5e^2}{4})\, K_{2p}(n) + \frac{81}{16}e^2\, K_{2p}(2n)\Bigg) \\
&\quad -ne\frac{m_p}{m_s}\left(\frac{R_s}{a}\right)^5 \Bigg(-(1-\frac{e^2}{4})\frac{3}{16}\, K_{2s}(n-2\Omega_s) \\
&\qquad -\frac{3}{4}(1-\frac{21e^2}{4})\, K_{2s}(2n-2\Omega_s) \\
&\qquad +\frac{147}{16}(1-\frac{179e^2}{28}\, K_{2s}(3n-2\Omega_s) \\
&\qquad +\frac{867}{8}e^2\, K_{2s}(4n-2\Omega_s) \\
&\qquad +\frac{9}{8}(1+\frac{5e^2}{4})\, K_{2s}(n) + \frac{81}{16}e^2\, K_{2p}(2n)\Bigg)
\end{aligned} \qquad (23)$$

where for any given frequency $\omega$ which can be either positive or negative, using the classical definition of $Q$ being definite positive, we have $K_{2p}(\omega) = k_{2p}(\omega) \times sign(\omega)/Q_p(\omega)$ and $K_{2s}(\omega) = k_{2s}(\omega) \times sign(\omega)/Q_s(\omega)$ where $k_2(\omega)$ and $Q_2(\omega)$ denotes the tidal Love number $k$ and tidal dissipation $Q$ for degree-2 tides calculated at the frequency $\omega$.

While both $k$ and $Q$ are frequency-dependent, this dependency is typically lower than 1% over a range of several hours (e.g., Pou et al., 2022). As this work is mainly an order-of-magnitude estimation, we assume here that both the Love number $k_2$ and the tidal dissipation factor $Q$ are independent of frequency, simplifying the change in eccentricity as:

$$\begin{aligned}
\dot{e} &= (\dot{e})_{primary} + (\dot{e})_{secondary} \\
&= -ne\frac{m_s}{m_p}\left(\frac{R_p}{a}\right)^5 \frac{k_{2p}}{Q_p}\Big(-(1-\frac{e^2}{4})\frac{3}{16}\,sign(n-2\Omega_p) \\
&\qquad -\frac{3}{4}(1-\frac{21e^2}{4})\,sign(2n-2\Omega_p) \\
&\qquad +\frac{147}{16}(1-\frac{179e^2}{28}\,sign(3n-2\Omega_p) \\
&\qquad +\frac{867}{8}e^2\,sign(4n-2\Omega_p) \\
&\qquad +\frac{9}{8}(1+\frac{5e^2}{4})+\frac{81}{16}e^2\Big) \\
&\quad -ne\frac{m_p}{m_s}\left(\frac{R_s}{a}\right)^5 \frac{k_{2s}}{Q_s}\Big(-(1-\frac{e^2}{4})\frac{3}{16}\,sign(n-2\Omega_s) \\
&\qquad -\frac{3}{4}(1-\frac{21e^2}{4})\,sign(2n-2\Omega_s) \\
&\qquad +\frac{147}{16}(1-\frac{179e^2}{28}\,sign(3n-2\Omega_s) \\
&\qquad +\frac{867}{8}e^2\,sign(4n-2\Omega_s) \\
&\qquad +\frac{9}{8}(1+\frac{5e^2}{4})+\frac{81}{16}e^2\Big)
\end{aligned} \qquad (24)$$

As shown in Eq. A.1, the change in eccentricity depends on the relative values of the mean motion $n$ and the bodies' spin period $\Omega$. Further simplification can be considered: from Table 1, we see that all primaries are significantly faster than the orbital periods of their secondaries, meaning that $\Omega_p$ is significantly greater than $n$. For the secondaries, as assumed in Murray and Dermott 2000 and shown on Fig. 1, most of the eccentricity damping occurs when the secondary is tidally locked. Thus we can assume $n = \Omega_s$, simplifying Eq. A.1 into Eq. 6.3:

$$\begin{aligned}
\dot{e} &= (\dot{e})_{primary} + (\dot{e})_{secondary} \\
&= -ne\frac{m_s}{m_p}\left(\frac{R_p}{a}\right)^5 \frac{k_{2p}}{Q_p}\left(-\frac{57}{8}-\frac{1509e^2}{32}\right) \\
&\quad -ne\frac{m_p}{m_s}\left(\frac{R_s}{a}\right)^5 \frac{k_{2s}}{Q_s}\left(\frac{21}{2}+\frac{903e^2}{4}\right)
\end{aligned} \qquad (25)$$

### A.2. Derivation of the spin evolution

From Equation (143) in Boué and Efroimsky (2019), the change in semi-major axis $a$ can be written to order 2 in inclination and order 4 in eccentricity as:

$$\dot{a} = (\dot{a})_{primary} + (\dot{a})_{secondary}$$

$$= \frac{m_s}{m_p} \left(\frac{R_p}{a}\right)^5 \left(-3an(1-5e^2) K_{2p}(2n-2\Omega_p)\right.$$
$$-\frac{9}{4}ane^2 K_{2p}(2n)$$
$$-\frac{3}{8}ane^2 K_{2p}(n-2\Omega_p)$$
$$\left.-\frac{441}{8}ane^2 K_{2p}(3n-2\Omega_p)\right) \quad (26)$$
$$-\frac{m_p}{m_s} \left(\frac{R_s}{a}\right)^5 \left(-3an(1-5e^2) K_{2s}(2n-2\Omega_s)\right.$$
$$-\frac{9}{4}ane^2 K_{2s}(2n)$$
$$-\frac{3}{8}ane^2 K_{2s}(n-2\Omega_s)$$
$$\left.-\frac{441}{8}ane^2 K_{2s}(3n-2\Omega_s)\right)$$

with the same notation as Eq. A.1. Assuming conservation of angular momentum in the system as done in Murray and Dermott (2000) and Boue and Efroimsky 2019, the relation between semi-major axis, primary spin, and secondary spin is given by:

$$\dot{\Omega}_p = -\frac{1}{2} \frac{1}{\alpha_p R_p^2} \frac{m_s}{m_p + m_s} na \, (\dot{a})_{primary} \quad (27)$$

$$\dot{\Omega}_s = -\frac{1}{2} \frac{1}{\alpha_s R_s^2} \frac{m_p}{m_p + m_s} na \, (\dot{a})_{secondary} \quad (28)$$

hence the following expression for change in the spin rate of primaries and secondaries:

$$\dot{\Omega}_p = -\frac{1}{2\alpha_p} \left(\frac{R_p}{a}\right)^3 \frac{m_s^2}{m_p(m_p + m_s)} n^2 \left(-3(1-5e^2) K_{2p}(2n-2\Omega_p)\right.$$
$$-\frac{9}{4}e^2 K_{2p}(2n)$$
$$-\frac{3}{8}e^2 K_{2p}(n-2\Omega_p) \quad (29)$$
$$\left.-\frac{441}{8}e^2 K_{2p}(3n-2\Omega_p)\right)$$

$$\dot{\Omega}_s = -\frac{1}{2\alpha_s} \left(\frac{R_s}{a}\right)^3 \frac{m_p^2}{m_s(m_s + m_p)} n^2 \left(-3(1-5e^2) K_{2s}(2n-2\Omega_s)\right.$$
$$-\frac{9}{4}e^2 K_{2s}(2n)$$
$$-\frac{3}{8}e^2 K_{2s}(n-2\Omega_s) \quad (30)$$
$$\left.-\frac{441}{8}e^2 K_{2s}(3n-2\Omega_s)\right)$$

The impact of high-order eccentricity terms on the spin evolution is trickier given that the assumption $n = \Omega_s$ cannot be made as for the eccentricity damping case. Instead, we choose to look at which initial value of $\Omega_s$ impacts the most our bounds based on the tidal locking timescale $\tau_\Omega$. $\tau_\Omega$.

Since $\tau_{\Omega_p}$ is used as a lower bound for primaries, the worst-case would be that the high-eccentricity terms are opposite to the order-0 term; but because we are working with fast-rotating primaries, we also choose to have $\Omega_p \geq 2n$. From this fast rotator assumption, Eq. A.2 can be simplified to:

$$\dot{\Omega}_p = -\frac{1}{2\alpha_p}\left(\frac{R_p}{a}\right)^3 \frac{m_s^2}{m_p(m_p + m_s)} n^2 \Bigg( +3(1 - 5e^2)\frac{k_{2p}(2n - 2\Omega_p)}{Q_p(2n - 2\Omega_p)} \\ - \frac{9}{4}e^2\frac{k_{2p}(2n)}{Q_p(2n)} \\ + \frac{3}{8}e^2\frac{k_{2p}(n - 2\Omega_p)}{Q_p(n - 2\Omega_p)} \\ + \frac{441}{8}e^2\frac{k_{2p}(3n - 2\Omega_p)}{Q_p(3n - 2\Omega_p)} \Bigg) \tag{31}$$

and keeping our previous assumption of having $k_2$ and $Q$ be frequency independent gives Eq. 6.3:

$$\dot{\Omega}_p = -\frac{1}{2}\frac{k_{2p}}{\alpha_p Q_p}\frac{m_s^2}{m_p(m_p + m_s)}\left(\frac{R_p}{a}\right)^3 n^2 \left(3 + \frac{153}{4}e^2\right) \tag{32}$$

Since $\tau_{\Omega_s}$ is used as an upper bound for the first secondary of asteroid (3749) Balam, the worst-case is for them to be of the same sign as the order-0 terms. Unlike their primaries, no other assumptions on their initial spin state can be made: therefore, the worst case for the evolution of $\Omega_s$ is when $\Omega_s \leq n/2$:

$$\dot{\Omega}_s = -\frac{1}{2\alpha_s}\left(\frac{R_s}{a}\right)^3 \frac{m_p^2}{m_s(m_s + m_p)} n^2 \Bigg( -3(1 - 5e^2)\frac{k_{2s}(2n - 2\Omega_s)}{Q_s(2n - 2\Omega_s)} \\ - \frac{9}{4}e^2\frac{k_{2s}(2n)}{Q_s(2n)} \\ - \frac{3}{8}e^2\frac{k_{2s}(2n - \Omega_s)}{Q_s(2n - \Omega_s)} \\ - \frac{441}{8}e^2\frac{k_{2s}(3n - 2\Omega_s)}{Q_s(3n - 2\Omega_s)} \Bigg) \tag{33}$$

and keeping our previous assumption of having $k_2$ and $Q$ be frequency independent gives Eq. 6.3:

$$\dot{\Omega}_s = -\frac{1}{2}\frac{k_{2s}}{\alpha_s Q_s}\frac{m_p^2}{m_s(m_s + m_p)}\left(\frac{R_s}{a}\right)^3 n^2 \left(-3 - \frac{171}{4}e^2\right) \tag{34}$$